\documentclass[preprint,12pt,numbers,sort&compress]{elsarticle}
\usepackage{amssymb}
\usepackage{graphicx}
\usepackage{float}
\usepackage{subfig}
\usepackage{epstopdf}
\usepackage{color} 
\usepackage{ae}
\journal{Physica A}

\begin{document}

\begin{frontmatter}
\title{Diffusion behavior of water confined in deformed carbon nanotubes}
\author[label1,label2]{Bruno H. S. Mendon\c{c}a\corref{cor1}}
\address[label1]{Instituto de F\'{i}sica, Universidade Federal do Rio Grande do Sul, Porto Alegre, 
RS 91501-970, Brazil} 
\address[label2]{Departamento de F\'{i}sica, Universidade Federal de Ouro Preto, Ouro Preto, MG 35400-000, Brazil}
\cortext[cor1]{Corresponding author.}
\ead{brunnohennrique13@gmail.com}
\author[label2,label3]{D\'{e}bora N. de Freitas}
\address[label3]{Departamento de F\'{i}sica, Universidade Federal de Juiz de Fora, 
Juiz de Fora, MG 36036-330, Brazil}
\author[label1]{Mateus H. K\"ohler}
\author[label2]{Ronaldo J. C. Batista}
\author[label1]{Marcia C. Barbosa}
\author[label2]{Alan B. de Oliveira}
\begin{abstract}
We use molecular dynamics simulations  to study the diffusion of water inside deformed carbon nanotubes
with different degrees of eccentricity at $300$K. We found a water structural transition between tubular-like 
to single-file for  (7,7) nanotubes associated with change from a high to  low mobility regimes.
Water  is frozen  when confined in a perfect (9,9) nanotube and it becomes liquid if such a nanotube is deformed above a certain 
threshold. Water diffusion enhancement (suppression)  is related to a reduction (increase) in the number 
of hydrogen bonds. This suggests that the shape of the nanotube is an  important ingredient when 
considering the dynamical and structural properties of confined water.
\end{abstract}
\begin{keyword}
Confined Water; Mobility; Carbon Nanotube; Diffusion.
\end{keyword}
\end{frontmatter}

%%%%%%%%%%%%%%%%%%%%%%%%%%%%%
\section{Introduction}
%%%%%%%%%%%%%%%%%%%%%%%%%%%%%%

Fluids under nanoscale confinement 
exhibit  properties not observed in the bulk \citep{HOL06,Bradley16}.
In the case of  water this shows an even larger
impact. When confined in
carbon nanotubes (CNTs) water  exhibits flow rates which exceeds
 by three orders of magnitude the values predicted by
the continuum hydrodynamic theory
\citep{majumder2005nanoscale,GMT10,NIC12,RIT14,LIU14,HOL14,RIT15,LIU16,MAT17,KONS15}. The superflow is
not the only anomalous behavior
observed in nanoconfined  water. It also presents
multi-phase  flow,
structural transitions
and highly heterogeneous hydrogen bonds
distribution \citep{nomura-pnas2017,farimani-jpcc2016,ternes2017single}.

For instance, the water diffusion coefficient 
in pristine carbon nanotubes does not decreases
 monotonically with the diameter of the tube \citep{barati2011}. Instead, it
has a  minimum  for the   (9,9) CNT, a maximum  for the (20,20) nanotube and
it approaches the bulk value for larger tubes.
The nanoconfined water forms layers and 
molecules 
near the wall diffuse faster than the particles in
the middle of the tube. This higher mobility
is caused by dangling hydrogen bonds at the
water-wall interface.

Although   pristine nanotubes
pose as  perfect models for studying  the confined water superflux  \citep{barati2011},
experimentally it is common to obtain nanotubes with defects,
vacancies and structural distortions \citep{sisto-cs2016,kroes-jctc2015}.
In addition functional groups may be adsorbed onto  tube's surface,
deposited at it's entrance
or even incorporated under compression.
All these factors lead to  structural deformations \citep{de2016vibrational,umeno2004theoretical},
which in turn can  affect the anomalous properties of the confined water.
For example,  the water streaming velocity and flow rate
 depend on the tube flexibility~\citep{sam-jcp2017} and
the effective shear stress and  viscosity  depend on the nanotube roughness,
which affects more  smaller tubes~\citep{xu-jcp2011}. These  observations were also
supported by experiments revealing  radius-dependent on the 
surface slippage in carbon nanotubes~\citep{secchi-nature2016},
and by simulations relating the shape of the nanotube with the
 dynamics of confined water 
with high influence on its flow and structure~\citep{belin-prf2016}.

Carbon nanotubes have been speculated to be present in virtually all areas of life and physical sciences  in a near future. From
drug delivery to water desalination, the existent literature  is vast. More specifically, several applications in nanofluidics have been explored. Examples include carbon nanotubes  as nanosyringes~\citep{HOL04} and nanothermometers~\citep{GAO02}. Studies focusing fluid transport  in carbon nanotubes are ubiquitous,  with interest in possible practical applications and also in 
water properties itself, when confined in such a peculiar media~\citep{HUM01,MAJ05,BRO01,WHI07}. 

In the real world carbon nanotubes are in the presence of substances, not only in its surroundings but filled with them. The contact between 
carbon nanotubes and substrates and/or surrounding adsorbates certainly change their structure. In this sense, the purpose of our work is to 
understand how deformations in carbon nanotubes change the diffusion of  the confined water. Considering the majority of literature regarding 
water diffusion in carbon nanotubes approach the problem using perfect nanotubes, we believe our work may fill a small piece of this important puzzle. 

Here we explore  in a systematic way
 by computer simulations  the effect of the change in
the eccentricity of the nanotube on the behavior of the diffusion
coefficient for different nanotube diameters. The idea behind
this work is to explore if there is a threshold distortion limit
beyond which the diffusion is non anomalous. This means that it would decrease
with the diameter of the nanotube as in  normal, 
non water-like fluids.  

The paper is organised as follows.
We present details of  simulations in Sec. \ref{methods},
in Sec. \ref{results} we discuss the results and  the Sec. \ref{conclusions} 
ends the paper with our conclusions.

%%%%%%%%%%%%%%%%%%%%%%%%%%%%%%%%%%%%%%%%%%%%%%%%%%%%%%%%%
\section{Computational Details and Methods}
\label{methods}
%%%%%%%%%%%%%%%%%%%%%%%%%%%%%%%%%%%%%%%%%%%%%%%%%%%%%%%%%

We performed molecular dynamics (MD) simulations for
the TIP4P/2005 water model \citep{abascal-jcp2005} using the 
LAMMPS package \citep{plimpton1995fast}. The  
SHAKE algorithm \citep{ryckaert1977numerical} was used to
 keep water molecules structure.
The choice of TIP4P/2005 over many other models available in literature
 was due to its accuracy in calculating water transport properties
at ambient conditions \citep{gonzalez-jcp2010}.
We represented the non-bonded interactions (carbon-oxygen) by the
 Lennard-Jones (LJ) potential with parameters 
$\epsilon_{CO}=0.11831$ kcal/mol and $\sigma_{CO}=3.28$ \AA ~\citep{HUM01}.
Interaction between carbon and hydrogen was set to zero.
 LJ cutoff distance was 12 \AA~ and long-range Coulomb interactions
 were treated using the particle$-$particle particle$-$mesh 
method. \citep{hockney1981} 
The time step used was 1 fs. The nanotubes considered were armchair
 with index n = 7, 9, 12, 16, 20, and 40. 

During carbon nanotubes deformation process,  carbon-carbon interaction
was modelled via  AIREBO 
potential \citep{stuart2000reactive,brenner2002second}. After
 reaching the desired degree of deformation, nanotubes were frozen, i.e., 
 carbon atoms neither interact  nor move relative to each other.

\begin{figure}[H]
\centering
\includegraphics[width=10.cm]{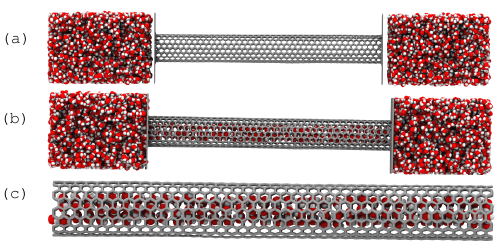}
\caption{The three main steps for filling nanotubes with water are as follows. (a) Undeformed carbon nanotubes are placed between water reservoirs at 300 K and 1 atm. (b) After a few nanoseconds they become completely filled with water. (c) The reservoirs are removed and the nanotubes are made periodic in  the axial direction.}
\label{abc}
\end{figure}

\begin{figure}[H]
\centering
\includegraphics[width=16.cm]{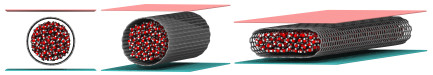}
\caption{Carbon nanotube between parallel plates.}
\label{fig-amass}
\end{figure}

Nanotubes filling process goes as follows.  First, water reservoirs
 containing a total of 8.000 water molecules were 
connected to both nanotubes ends as shown in  Figure \ref{abc}(a). Pressure 
and temperature of reservoirs were kept at 1 atm and 300 K, respectively, by
 means of Nos{\'e}-Hoover barostat and thermostat. After
a few nanoseconds the  equilibrium configuration in which  nanotubes  are filled with water
is reached  as depicted in  Figure \ref{abc}(b). Next, the reservoirs
 are removed and the simulation box size is adjusted to fit the
 nanotubes length as shown in  Figure \ref{abc}(c). Boundary conditions
 were periodic in all directions during all stages. The number 
of water molecules enclosed depend on nanotubes dimensions. This information can be found in Table \ref{tab_CNT}.
%%%%%%%%%%%%%%%%%%%%%%%%%%%%%%%%%%%%%%%%%%%%%%%%%%%%%%%%%%%%%%%
%%%%%%%%%%%%%%%%%%%%%%%%%%%%%%%%%%%%%%%%%%%%%%%%%%%%%%%%%%%%%%
%%%%%%%%%%%%%%%%%%%%%%%%%%%%%%%%%%%%%%%%%%%%%%%%%%%%%%%%%%%%%
\begin{table}[H]
\begin{center}
  \caption{Carbon nanotubes  diameter $d$, length $L_z,$ and  and number of water molecules inside.} 
  \begin{tabular}{ c c c c c c }
    \hline \hline
    CNT (n,n) $\quad$ & $\quad$ $d$ (nm) & $\quad$ $L_{z}$ (nm) & $\quad$ Number of $H_{2}$O    \\ \hline
    (7,7)     $\quad$ & $\quad$ 0.95	 & $\quad$ 123.4         & $\quad$ 901                                  \\ 
    (9,9)     $\quad$ & $\quad$ 1.22	 & $\quad$ 50.66         & $\quad$ 908                                  \\ 
    (12,12)   $\quad$ & $\quad$ 1.62	 & $\quad$ 22.62         & $\quad$ 901                                  \\ 
    (16,16)   $\quad$ & $\quad$ 2.17	 & $\quad$ 11.06         & $\quad$ 911                                  \\ 
    (20,20)   $\quad$ & $\quad$ 2.71	 & $\quad$ 10.33         & $\quad$ 1440                                  \\ 
    (40,40)   $\quad$ & $\quad$ 5.42	 & $\quad$ 7.87          & $\quad$ 5221                                  
    \\ \hline \hline
  \end{tabular}
\label{tab_CNT}
\end{center}
\end{table}
%%%%%%%%%%%%%%%%%%%%%%%%%%%%%%%%%%%%%%%%%%%%%%%%%%%%%%%%%%%%%%%%%%%%%%%
After  perfect nanotubes are filled with water, they
 are compressed to different degrees of deformation by loading them with plates made of 
frozen atoms (see  Figure \ref{fig-amass}).  
 The interaction between  CNTs and plates was given by the repulsive part of LJ potential with 
  energy parameter $\epsilon =  0.184$ kcal/mol and distance  parameter $\sigma = 3.0$ \AA. 
 The cutoff used was 3.0 \AA. 
For compressing 
the nanotubes 
plates are approached to each other at constant speed
 until the nanotube reaches the desired  deformation.  Speeds 
were in the range from 0.2 to 0.6 \AA/ps.

%%%%%%%%%%%%%%%%%%%%%%%%%%%%%%%%%%%%%%%%%%%%%%%%%%%%%%%%%%%%%%%%%
\begin{figure}[H]
\centering
\includegraphics[width=13.5cm]{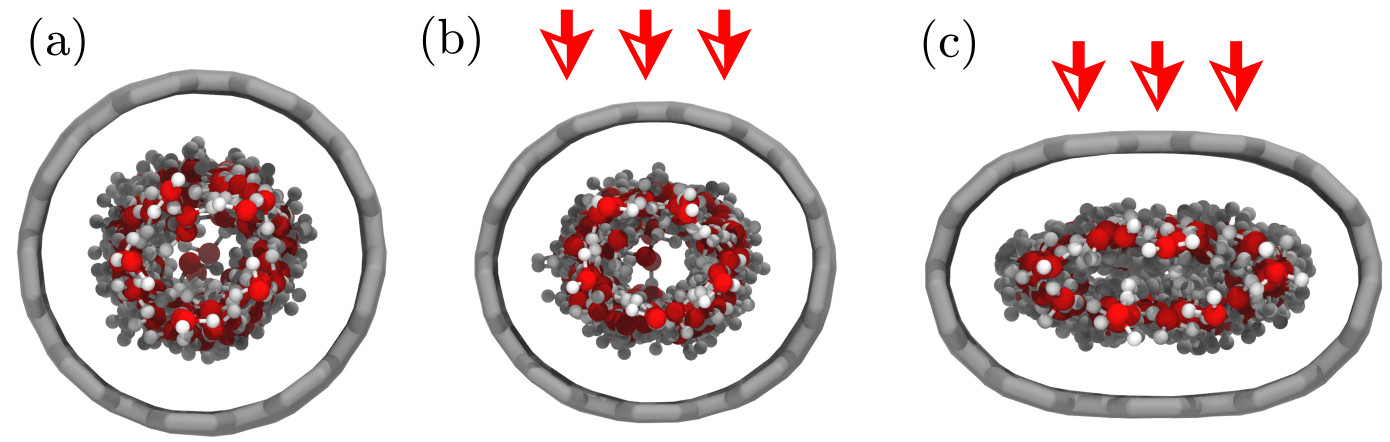}
\caption{Snapshots of the (9,9) CNT at (a) $e=0.0,$ (b) $e=0.4,$ and (c) $e=0.8$.
}
\label{fig1}
\end{figure}

%%%%%%%%%%%%%%%%%%%%%%%%%%%%%%%%%%%%%%%%%%%%%%%%%%%%%%%%%%%%%%%%
Nanotubes deformation are characterised by eccentricity after compression, which reads
%%%%%%%%%%%%%%%%%%%%%%%%%%%%%%%%%%%%%%%%%%%%%%%%%%%%%%%%%%%%%%%%
\begin{equation}
\label{eq:dist0029}
e= \sqrt{1 - \frac{b^{2}}{a^{2}}},
\end{equation}
%%%%%%%%%%%%%%%%%%%%%%%%%%%%%%%%%%%%%%%%%%%%%%%%%%
\noindent where $b$ and $a$ being the smaller and larger 
semi-axis, respectively. 
We approached  nanotubes with eccentricities
ranging from 0.0 (perfect) to 0.8 (highly deformed) as 
shown in  Figure~\ref{fig1}.

Then, long 
simulations of water inside  deformed carbon nanotubes
were carried out for nanotubes with different diameters. 
The production part was conducted  in the canonical 
ensemble, with temperature fixed at 300 K
using the Nos\'{e}-Hoover thermostat \citep{nose1984unified}
with a time constant of 0.1 ps. After filling and deforming 
nanotubes, the system was allowed to 
equilibrate for 5 ns before data collection.
%%%%%%%%%%%%%%%%%%%%%%%%%%%%%%%%%%%%%%%%%%%%%%%%%%%%%
%%%%%%%%%%%%%%%%%%%%%%%%%%%%%%%%%%%%%%%%%%%%%%%%%%%%%%%%
To study the mobility we employed the
 mean squared displacement (MSD) given by: \citep{striolo2006mechanism}
%%%%%%%%%%%%%%%%%%%%%%%%%%%%%%%%%%%%%%%%%%%%%%%%%%%%%%%
\begin{eqnarray}
\label{eq1}
\langle \Delta 
\vec{r}(t)^2 \rangle=\left \langle  \left | \vec{r}(t)-\vec{r}(0) \right |^{2} \right \rangle 
\end{eqnarray}
%%%%%%%%%%%%%%%%%%%%%%%%%%%%%%%%%%%%%%%%%%%%%%%%%%%%%%%%%
where $\left \langle  \left | \vec{r}(t)-\vec{r}(0) \right |^{2} \right \rangle$ is referred as the MSD,
$\left \langle \right \rangle$ denotes an average
over all  molecules
and $\vec{r}\left ( t \right )$ is the displacement
of a molecule during the time interval $t$.
Diffusion constant $D$ is related to MSD and time through the relation
%%%%%%%%%%%%%%%%%%%%%%%%%%%%%%%%%%%%%%%%%%%%%%%%%%%%%%%%%
\begin{eqnarray}
\langle \Delta 
\vec{r}(t)^2 \rangle\propto D t^{\alpha},
\label{diffusionalpha}
\end{eqnarray}
%%%%%%%%%%%%%%%%%%%%%%%%%%%%%%%%%%%%%%%%%%%%%%%%%%%%%%%%%%%
\noindent where  $\alpha$ is a signature of which type of diffusive 
regime the system is following namely
$\alpha=$ 1 is the Fickian diffusion,
$\alpha>$ 1 indicates the superdiffusive regime and  $\alpha<$ 1 
refers to the sub-diffusive regime.
In the bulk phase,
 water molecules obey  Fickian diffusion.
When confined in CNTs
the diffusion  of water molecules becomes more involving due to the
nanoscale confinement. \citep{striolo2006mechanism}

For the hydrogen bonds statistics  we used the geometrical criteria
of donor-hydrogen-acceptor (DHA) angle and donor-acceptor (DA)  
distance. A hydrogen bond is computed if DHA angle $\leq$ 30$^{\circ}$ 
and  DA distance < 0.35 nm  \citep{spoel-jpcb2006,joseph2008}. 
%%%%%%%%%%%%%%%%%%%%%%%%%%%%%%%%%%%%%%%%%%%%%%%%%%%%%%%%%%%%
\section{Results}
\label{results}
\begin{figure}[H]
\begin{center}
\includegraphics[width=10.5cm]{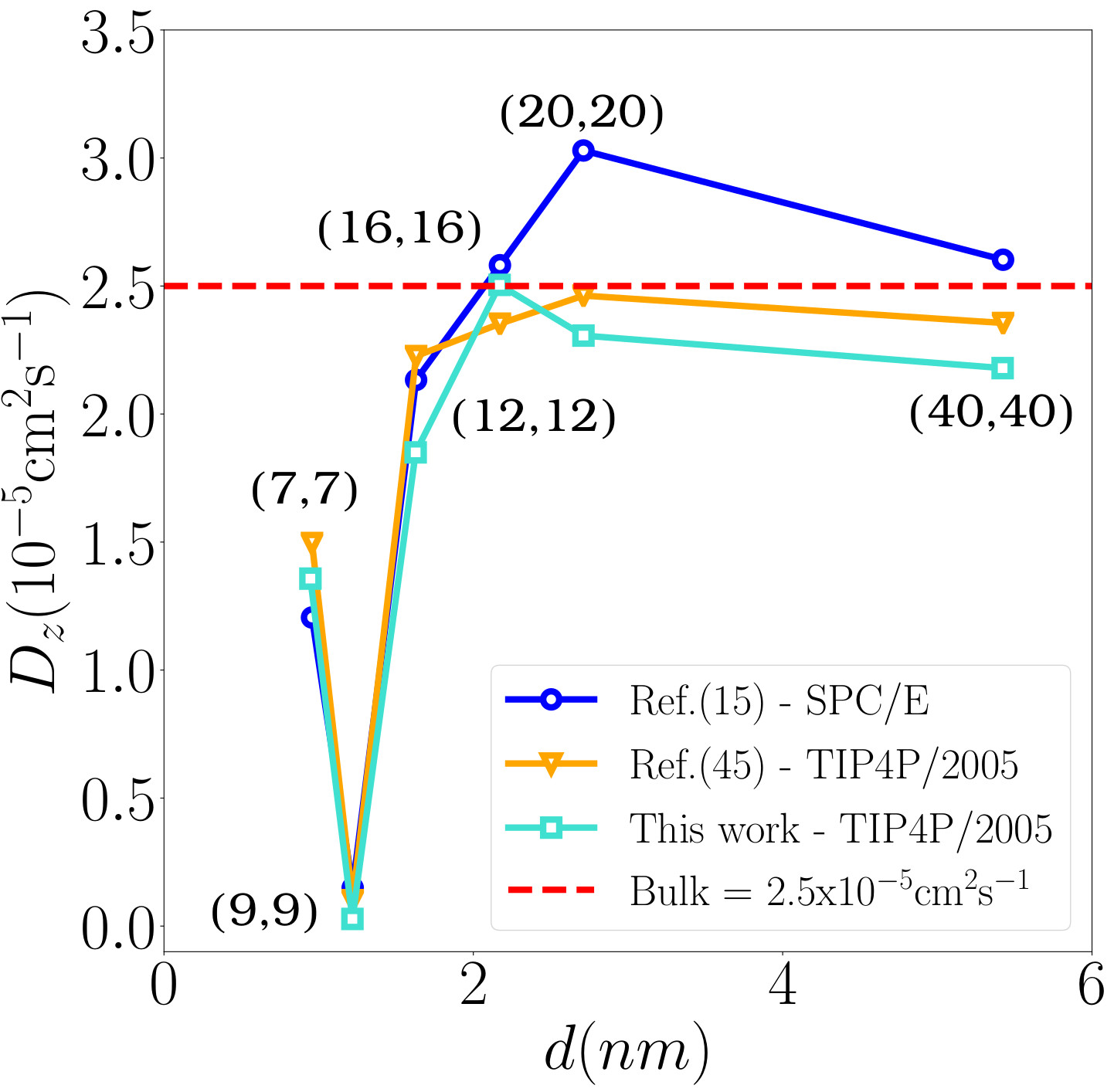}
\caption{\label{fig:fig-diff-ref}Diffusion coefficient versus diameter of the  carbon nanotube for the perfect $e=$0 tube .}
\end{center}
\end{figure}

%%%%%%%%%%%%%%%%%%%%%%%%%%%%%%%%%%%%%%%%%%%%%%%%%%%%%%%%%%%
In  Figure~\ref{fig:fig-diff-ref}
we show  the diffusion coefficient of water as a
 function of the diameter for the perfect nanotube ( $e = 0$)
for the  TIP4P/2005 water model (both from this work and from  K\"{o}hler et al. \citep{kohler2016size}) and for the SPC/E water from Ref. \citep{barati2011}.
For the two models, the
diffusion coefficient shows
 a global minimum at  1 nm diameter which corresponds to a (9,9) CNT.
In this case the mobility of  molecules
is virtually zero. The water is  frozen inside the nanotube
assuming a solid  ring-like structure,
as can be seen in Figure~\ref{fig3}(c). This ring is immobile and each molecule
makes a large number of hydrogen bonds as shown in  Figure~\ref{fig5}.
The low
dynamics  seen in water confined in (9,9) CNT is due to the
commensurability  with the hydrogen bonds distance and (9,9) CNT diameter,
which favours the formation of an organised network as
 Figure~\ref{fig3} illustrates.
The interplay between the diffusion coefficient showed
in  Figure~\ref{fig3}
and the hydrogen bonds network illustrated 
in  Figure~\ref{fig5} is the mechanism
behind the anomalous behaviour
 of confined water \citep{kohler2016size,barati2011}. 
%%%%%%%%%%%%%%%%%%%%%%%%%%%%%%%%%%%%%%%%%%%%%%%%%%%%%%
%%%%%%%%%%%%%%%%%%%%%%%%%%%%%%%%%%%%%%%%%%%%%%%%%%%
For larger nanotube diameters the diffusion
coefficient approaches the bulk value
(around 2.5$\times10^{-5}$ cm$^2$/s) \citep{harris1980pressure}.
For intermediate nanotube diameters (2-3 nm),
there is a maximum in the diffusion coefficient.
Larger surface areas induce more 
dangling bonds,
while  large central  volumes 
favours hydrogen bonds formation. The
minimum and  maximum observed in the 
diffusion coefficient are related to this 
competition between the water-wall contact area
and the volume occupied by the fluid \citep{barati2011}.

\begin{figure}[H]
\centering
\includegraphics[width=15.5cm]{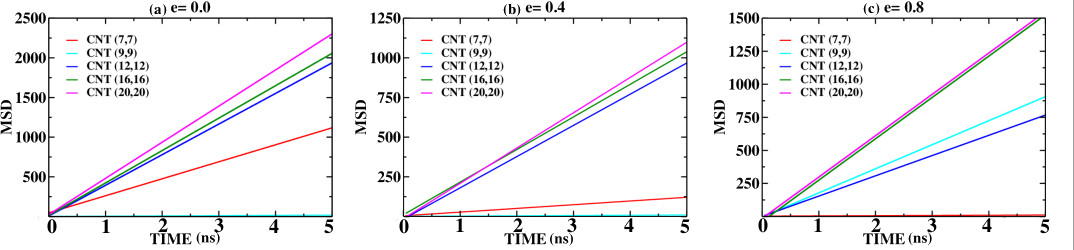}
\caption{MSD curves as a function of time for the carbon nanotubes n= 7, 9, 12, 16 and 20 for
(a) $e=$ 0.0,
(b) $e=$ 0.4 and
(c) $e=$ 0.8.
}
\label{msd}
\end{figure}

\begin{figure}[H]
\centering
\includegraphics[width=15.5cm]{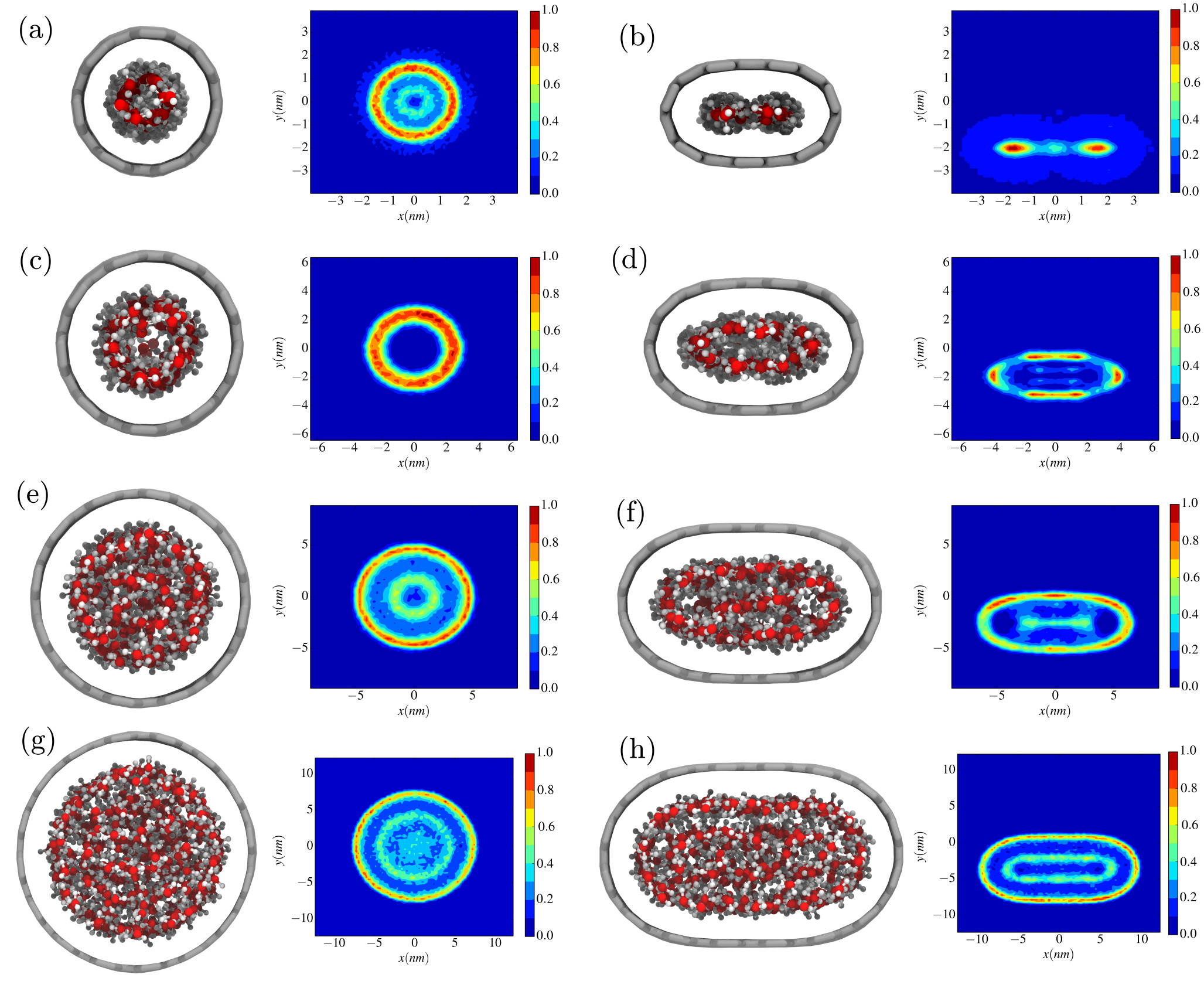}
\caption{Left panels: snapshots of water molecules inside nanotubes  along with radial ($x$-$y$) density maps of oxygens inside
(a) (7,7),
(c) (9,9),
(e) (12,12)
and (g) (16,16)
carbon nanotubes with eccentricity $e=$ 0.0.
Right panels show the correspondent nanotubes with $e=$ 0.8.}
\label{fig3}
\end{figure}

\begin{figure}[H]
\centering
\includegraphics[width=15.5cm]{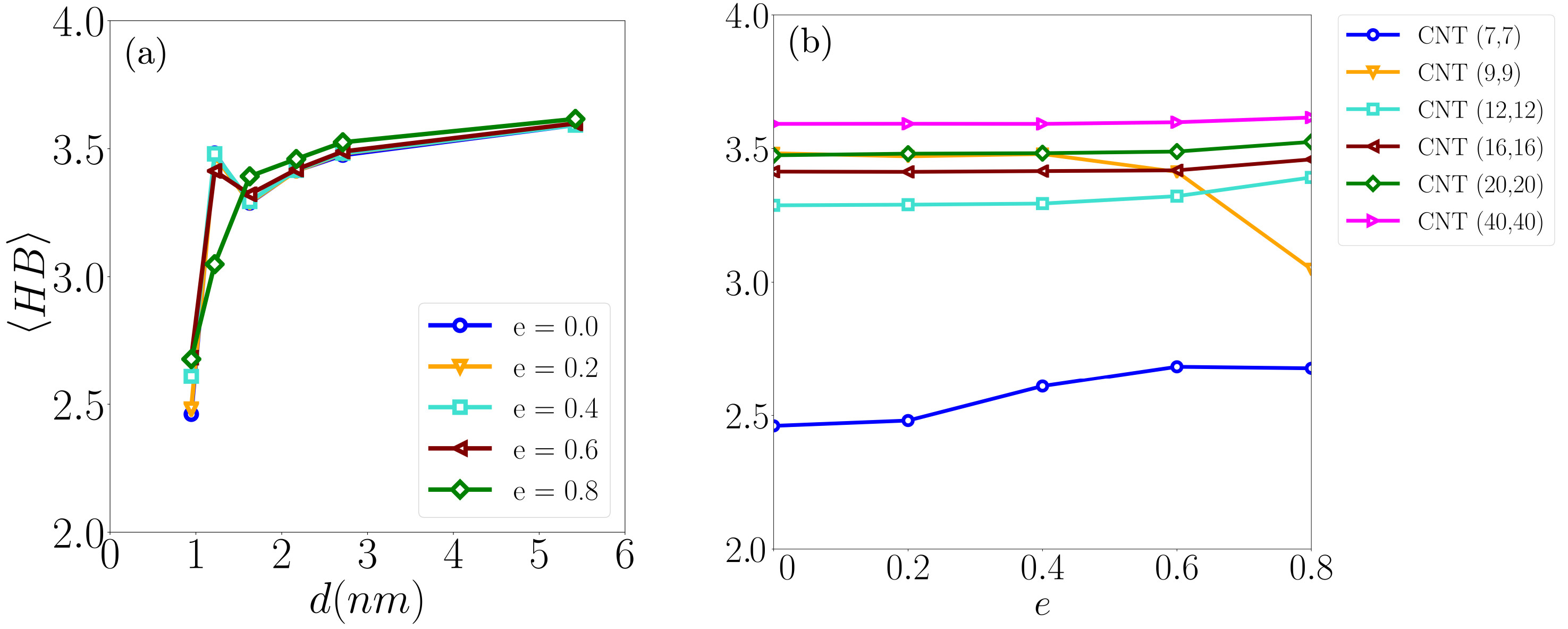}
\caption{Average number of hydrogen bonds (HB) of each  water molecule as a function of (a) the nanotube diameter relative of the nanotube $e=0.0$ and
(b) the eccentricity.}
\label{fig5}
\end{figure}

 In Eq. (\ref{diffusionalpha}) we found $\alpha=1$, which is a signature of Fickian diffusion. In  Fig. \ref{msd} we show  MSD curves as a function of time for  carbon nanotubes with n = 7, 9, 12, 16 and 20, with eccentricities (a) e = 0.0, (b) e= 0.4 and (c) e = 0.8.

%%%%%%%%%%%%%%%%%%%%%%%%%%%%%%%%%%%%%%%%%%%%%%%%%%%%%%%%%%
%%%%%%%%%%%%%%%%%%%%%%%%%%%%%%%%%%%%%%%%%%%%%%%%%%%%%%%%%%%%

Figure~\ref{fig3} shows snapshots of the last simulation steps
and the radial density maps for  the  nanotubes
with $e= 0.0$ (left) and $e= 0.8$ (right).
This  Figure also shows the oxygen density
maps  constructed by dividing the nanotubes radial direction  
into small concentric bins
and by averaging the number of oxygen atoms in each bin.
Red regions represent high probability of finding a water molecule
 while dark blue stands for low probability.
For the  (7,7) nanotube, [Figure~\ref{fig3}(a) and (b)] the increase
in  deformation  makes
 water molecules to undergo a structural transition:
from a cylindrical organisation to two single-file structures.
Molecules in one single-file tend to bond to particles in 
the other single-file.Then, the deformation in (7,7) nanotubes leads to a high-mobility to low-mobility transition
due to an increasing in the number of hydrogen bonds.
%%%%%%%%%%%%%%%%%%%%%%%%%%%%%%%%%%%%%%%%%%%%%%%%%%%%%%%%%%
%%%%%%%%%%%%%%%%%%%%%%%%%%%%%%%%%%%%%%%%%%%%%%%%%%%%%%%%%%%%

%%%%%%%%%%%%%%%%%%%%%%%%%%%%%%%%%%%%%%%%%%%%%%%%%%%%%%%%
%%%%%%%%%%%%%%%%%%%%%%%%%%%%%%%%%%%%%%%%%%%%%%%%%%%%%%%%
For the (9,9) nanotube [Figure~\ref{fig3}(c) and (d)]
as the deformation is imposed we observe 
a transition from  an uniform frozen
water layer to  non uniform liquid water layered structure.
For  $e=0.8$ clusters of fluid water at the wall 
replace the frozen $e=0.0$ structure. The deformation in the case (9,9)
leads to low-mobility for a high-mobility transition.

For the larger tubes,
(12,12) and (16,16),
[Figures~\ref{fig3}(e) to (h)] the
systems form
liquid layers, which
despite deformed are preserved under tension.
For (20,20) and (40,40) CNTs
no representative transformation is observed as we increase the eccentricity.
  
\begin{figure}[H]
\centering
\includegraphics[width=15.5cm]{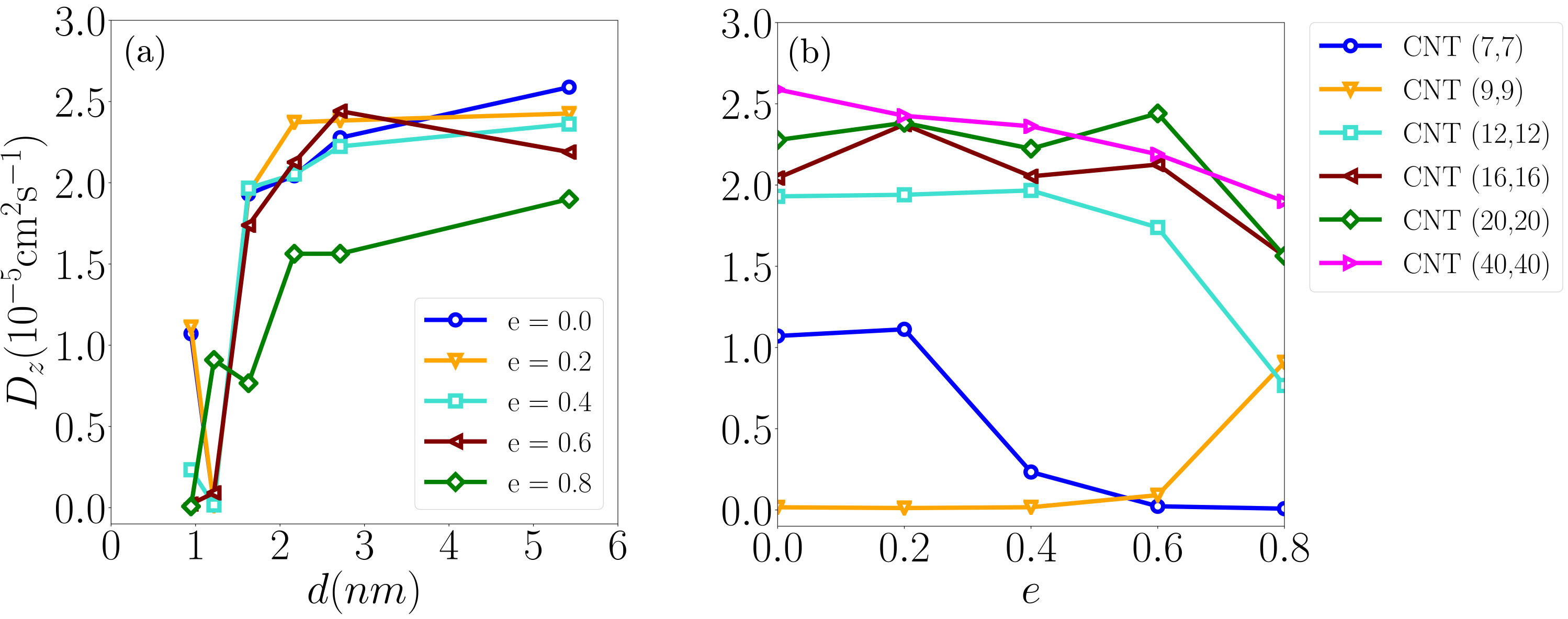}
\caption{Axial diffusion coefficient of water as a function of (a) the diameter relative of the nanotube $e=0.0$ and
(b) eccentricity.}
\label{fig4}
\end{figure}

In order to test if the structural transitions observed for the 
(7,7) and (9,9) nanotubes are related to 
changes in the mobility  MSDs of confined water
were computed. 
Figure~\ref{fig4}(a) shows
the diffusion coefficient of water as a function of the 
nanotube diameter $d$ for different eccentricities ($0 \leq e \leq 0.8$).
For the highest deformations ($e=0.6$ and $e=0.8$)
 systems show no increase in  $D$ for 
 diameters below 1.5 nm as observed for
the undeformed case. This suggests that the
deformation suppresses the anomalous high diffusion
(and maybe flux) observed in confined water.

Figure~\ref{fig4}(b)
illustrates the change in the water diffusion 
coefficient with deformation $e$.
For the (7,7) case the increase of $e$ leads
to a decrease in the diffusion coefficient, which 
is related to the structural change observed in the 
Figure~\ref{fig3}. This result indicates that any
amount of deformation destroys the super diffusion 
observed in confined water.
For the (9,9) case, the opposite happens. As 
the deformation goes beyond a certain threshold, the 
water frozen at the wall melts. However in 
this case the transition requires a huge deformation.
%%%%%%%%%%%%%%%%%%%%%%%%%%%%%%%%%%%%%%%%%%%%%
%%%%%%%%%%%%%%%%%%%%%%%%%%%%%%%%%%%%%%%%%%%%%%%%%%%%%%%%
For understanding the relation between dynamics 
and structure assumed by the confined water,
we calculated the average number of hydrogen bonds of each molecule 
for the different systems studied.
Figure.~\ref{fig5}(a)
shows that 
this number is highly dependent on the nanotube size
and tends to the bulk value for the larger nanotubes.
For the $e<0.8$ a local maximum in the number
of bonds occur at the (9,9) nanotube coinciding
with the minimum of diffusion.
For the $e=0.8$ no maximum in the number of bonds
is observed what is agreement with the non zero diffusion 
and with our suggestion
that the (9,9) system does not melt for the high deformation.

Figure~\ref{fig5} (b) illustrates
 the average number of hydrogen bonds of each water molecule
 as a function 
of the nanotube eccentricity.
For the 
the (7,7) nanotube
the water shows monotonically increase of hydrogen bonds
what supports the idea that the decrease in the mobility with
deformation is followed by the formation of a more bonded system. For the 
 (9,9) CNT, the reduction of the number of hydrogen bonds
at $e=0.8$ what supports the absence of melting in this case.

\begin{figure}[H]
\centering
\includegraphics[width=15.5cm]{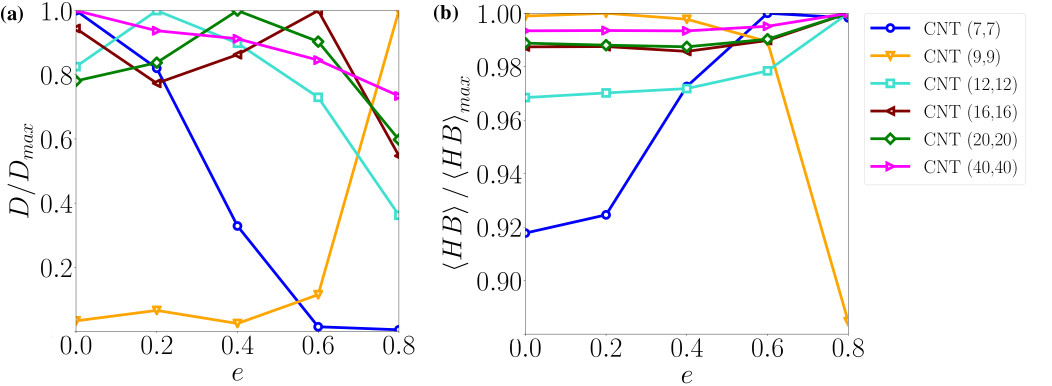}
\caption{Percentage fraction (a) D/D$_{max}$ and (b) <HB>/<HB>$_{max}$ per water molecule as a function of the nanotube eccentricity.}
\label{fig6}
\end{figure}

In order to confirm the correlation
between the behaviour of the mobility and 
the number of hydrogen bonds we compute
the percentual change of each of these quantities as follows.
For each nanotube size the diffusion and 
 number of hydrogen bonds  is renormalised by its
maximum value. 
Figure~\ref{fig6} illustrates the percentual
of the diffusion and of the number of bonds
D/D$_{max}$  and <HB>/<HB>$_{max}$ as a function
of the eccentricity. This figure confirms that
diffusion correlates with the number of 
hydrogen bonds. For the (7,7), (12,12), (16,16), (20,20) and (40,40)
the percentual diffusion decreases and the percentual number of
bonds decrease with 
the increase of the eccentricity while for 
the (9,9) case the system is frozen
for $e\leq 0.6$. This implies
that the melting transition when
the system changes from (9,9) to (7,7) is 
suppressed by the high eccentricity.

%%%%%%%%%%%%%%%%%%%%%%%%%%%%%%%%%%%%%%%%%%%%%%%%%%%%
%%%%%%%%%%%%%%%%%%%%%%%%%%%%%%%%%%%%%%%%%%%%%%%
\section{Conclusion}
\label{conclusions}
%%%%%%%%%%%%%%%%%%%%%%%%%%%%%%%%%%%%%%%%%%%%%%%%
We investigated the effect of deforming CNTs in the structure 
and diffusion of the confined water.

We show that the deformation of the 
nanotube suppresses two phenomena observed in
the diffusion of confined water: the 
increase of the diffusion with the decrease
of the nanotube diameter and the freezing of confined  water
above the melting transition. 

The disappearance of the diffusion enhancement
is observed because the  water at the (7,7) tube shows
a smooth  transition from a fluid to a frozen state as the 
tube is deformed.
The suppression of the frozen state for the water at 
the  (9,9) nanotube is observed by the increase of the 
water mobility at the deformation $e=0.8$.
In this case there is a critical diameter ($\sim$ 1-1.2 nm)
for perfect nanotubes inside which the water freezes.
By deforming the tube,
varying the eccentricity $e$ from 0 to 0.8,
we have shown that inside this pores
the hydrogen bonds break down
and the crystalline structure is no longer favourable,
leading to anomalous water diffusion.
As we increase the nanotube diameter,
the water diffusion presents a maximum in relation to the tube eccentricity,
but at this point the effect of deformation is less prominent.
This maximum is a consequence of
the competition between nanotube area and water volume:
the increase of the internal area of the tubes favours the break of hydrogen bonds in water (generating ``dangling bonds''),
while the increase in water volume favours the formation of hydrogen bonds.

For almost all systems,
the increase in nanotube deformation leads to increase the number of hydrogen bonds,
which slows down the water mobility.
The only exception is the CNT (9,9)
in which the deformation decreases the number of hydrogen bonds
and increase the water diffusion.
In this nanotube
the symmetrical shape leads to frozen water molecules at interface,
while the deformation leads to enhanced diffusion.
All of these evidences
point to the importance of the nanotube structure on the properties of the confined water.

\section*{Acknowledgments}
This work is partially supported by Brazilian agencies CNPq and CAPES, Universidade Federal de Ouro Preto, Universidade Federal 
do Rio Grande do Sul and INCT-Fcx. ABO thanks the Brazilian science agency FAPEMIG
for financial support through the Pesquisador 
Mineiro grant. BHSM is indebted to Patricia Ternes who inspired the graphics of work.  

\bibliographystyle{elsarticle-num}
\bibliography{sample}

\begin{thebibliography}{10}
\expandafter\ifx\csname url\endcsname\relax
  \def\url#1{\texttt{#1}}\fi
\expandafter\ifx\csname urlprefix\endcsname\relax\def\urlprefix{URL }\fi
\expandafter\ifx\csname href\endcsname\relax
  \def\href#1#2{#2} \def\path#1{#1}\fi

\bibitem{HOL06}
J.~K. Holt, H.~G. Park, Y.~Wang, M.~Stadermann, A.~B. Artyukhin, C.~P.
  Grigoropoulos, A.~Noy, O.~Bakajin, Science 312 (2006) 1034.

\bibitem{Bradley16}
J.~L. Bradley-Shaw, P.~J. Camp, P.~J. Dowding, K.~Lewtas, Langmuir 32 (2016)
  7707--7718.

\bibitem{majumder2005nanoscale}
M.~Majumder, N.~Chopra, R.~Andrews, B.~J. Hinds, Nature 438 (2005) 44.

\bibitem{GMT10}
G.~Torrie, G.~Lakatos, G.~Patey, The Journal of Chemical Physics 133 (2010)
  224703.

\bibitem{NIC12}
W.~Nicholls, M.~K. Borg, D.~A. Lockerby, J.~Reese, Molecular Simulation 38
  (2012) 781--785.

\bibitem{RIT14}
K.~Ritos, D.~Mattia, F.~Calabr{\`o}, J.~M. Reese, The Journal of Chemical
  Physics 140 (2014) 014702.

\bibitem{LIU14}
L.~Liu, G.~Patey, The Journal of Chemical Physics 141 (2014) 18C518.

\bibitem{HOL14}
D.~M. Holland, D.~A. Lockerby, M.~K. Borg, W.~D. Nicholls, J.~M. Reese,
  Microfluidics and Nanofluidics 18 (2015) 461--474.

\bibitem{RIT15}
K.~Ritos, M.~K. Borg, N.~J. Mottram, J.~M. Reese, Philosophical Transactions of
  the Royal Society A 374 (2016) 20150025.

\bibitem{LIU16}
L.~Liu, G.~Patey, The Journal of Chemical Physics 144 (2016) 184502.

\bibitem{MAT17}
M.~K. Borg, J.~M. Reese, MRS Bulletin 42 (2017) 294--299.

\bibitem{KONS15}
K.~Ritos, M.~K. Borg, D.~A. Lockerby, D.~R. Emerson, J.~M. Reese, Microfluidics
  and Nanofluidics 19 (2015) 997--1010.

\bibitem{nomura-pnas2017}
K.~Nomura, T.~Kaneko, J.~Bai, J.~S. Francisco, K.~Yasuoka, X.~C. Zeng,
  Proceedings of the National Academy of Sciences 114 (2017) 4066.

\bibitem{farimani-jpcc2016}
A.~Barati~Farimani, N.~R. Aluru, The Journal of Physical Chemistry C 120 (2016)
  23763.

\bibitem{ternes2017single}
P.~Ternes, A.~Mendoza-Coto, E.~Salcedo, The Journal of Chemical Physics 147
  (2017) 034510.

\bibitem{barati2011}
A.~Barati~Farimani, N.~Aluru, The Journal of Chemical Physics B 115 (2011)
  12145.

\bibitem{sisto-cs2016}
T.~J. Sisto, L.~N. Zakharov, B.~M. White, R.~Jasti, Chemical Science 7 (2016)
  3681.

\bibitem{kroes-jctc2015}
J.~M.~H. Kroes, F.~Pietrucci, A.~C.~T. van Duin, W.~Andreoni, Journal of
  Chemical Theory and Computation 11 (2015) 3393.

\bibitem{de2016vibrational}
A.~B. de~Oliveira, H.~Chacham, J.~S. Soares, T.~M. Manhabosco, H.~F.
  de~Resende, R.~J. Batista, Carbon 96 (2016) 616.

\bibitem{umeno2004theoretical}
Y.~Umeno, T.~Kitamura, A.~Kushima, Computational Materials Science 30 (2004)
  283.

\bibitem{sam-jcp2017}
A.~Sam, S.~K. Kannam, R.~Hartkamp, S.~P. Sathian, The Journal of Chemical
  Physics 146 (2017) 234701.

\bibitem{xu-jcp2011}
B.~Xu, Y.~Li, T.~Park, , X.~Chen, The Journal of Chemical Physics 135 (2011)
  144703.

\bibitem{secchi-nature2016}
E.~Secchi, S.~Marbach, A.~Nigues, D.~Stein, A.~Siria, L.~Bocquet, Nature 537
  (2016) 210.

\bibitem{belin-prf2016}
C.~Belin, L.~Joly, F.~Detcheverry, Physical Review Fluids 1 (2016) 054103.

\bibitem{HOL04}
J.~K. Holt, A.~Noy, T.~Huser, D.~Eaglesham, O.~Bakajin, Nano Letters 4 (2004)
  2245--2250.

\bibitem{GAO02}
Y.~Gao, Y.~Bando, Nature 415 (2002) 599.

\bibitem{HUM01}
G.~Hummer, J.~C. Rasaiah, J.~P. Noworyta, Nature 414 (2001) 188.

\bibitem{MAJ05}
M.~Majumder, N.~Chopra, R.~Andrews, B.~J. Hinds, Nature 438 (2005) 44.

\bibitem{BRO01}
I.~Brovchenko, A.~Geiger, A.~Oleinikova, Physical Chemistry Chemical Physics 3
  (2001) 1567--1569.

\bibitem{WHI07}
M.~Whitby, N.~Quirke, Nature Nanotechnology 2 (2007) 87.

\bibitem{abascal-jcp2005}
J.~L.~F. Abascal, C.~Vega, The Journal of Chemical Physics 123 (2005) 234505.

\bibitem{plimpton1995fast}
S.~Plimpton, Journal of Computational Physics 117 (1995) 1.

\bibitem{ryckaert1977numerical}
J.-P. Ryckaert, G.~Ciccotti, H.~J. Berendsen, Journal of Computational Physics
  23 (1977) 327.

\bibitem{gonzalez-jcp2010}
M.~Gonzalez, J.~Abascal, The Journal of Chemical Physics 132 (2010) 096101.

\bibitem{hockney1981}
R.~W. Hockney, J.~W. Eastwood, McGraw-Hill 1.

\bibitem{stuart2000reactive}
S.~J. Stuart, A.~B. Tutein, J.~A. Harrison, The Journal of Chemical Physics 112
  (2000) 6472.

\bibitem{brenner2002second}
D.~W. Brenner, O.~A. Shenderova, J.~A. Harrison, S.~J. Stuart, B.~Ni, S.~B.
  Sinnott, Journal of Physics: Condensed Matter 14 (2002) 783.

\bibitem{nose1984unified}
S.~Nos{\'e}, The Journal of Chemical Physics 81 (1984) 511.

\bibitem{striolo2006mechanism}
A.~Striolo, Nano Letters 6 (2006) 633.

\bibitem{spoel-jpcb2006}
D.~van der Spoel~et al., The Journal of Physical Chemistry B 110 (2006) 4393.

\bibitem{joseph2008}
S.~Joseph, N.~Aluru, Nano Letters 8 (2008) 452.

\bibitem{kohler2016size}
M.~H. K{\"o}hler, L.~B. da~Silva, Chemical Physics Letters 645 (2016) 38.

\bibitem{harris1980pressure}
K.~R. Harris, L.~A. Woolf, The Journal of the Chemical Society, Faraday
  Transactions 76 (1980) 377.

\end{thebibliography}
{\color{blue}

\end{document}